\documentclass[draft]{agujournal2019}
\usepackage{url} 
\usepackage{lineno}
\usepackage[inline]{trackchanges} 
\usepackage{soul}
\draftfalse

\journalname{Radio Science}

\begin{document}

%
%


\title{The impacts of tropospheric gravity wave-generated MSTIDs on skywaves at middle latitude North American sector observed and modeled using SuperDARN HF radars}

%
%




\authors{S. Chakraborty\affil{1}, P.A. Inchin\affil{2}, S. Debchoudhury\affil{1}, C. Heale\affil{1}, B. Bergsson\affil{1}, M. Zettergren\affil{1}, J. M. Ruohoniemi\affil{3}}


\affiliation{1}{Embry-Riddle Aeronautical University, Daytona Beach, FL, USA}
\affiliation{2}{Computational Physics Inc., VA, USA}
\affiliation{3}{Center for Space Science and Engineering Research, Virginia Tech, Blacksburg, VA, USA}




\correspondingauthor{Shibaji Chakraborty}{shibaji7@vt.edu}




\begin{keypoints}
\item Gravity wave-driven MSTIDs can be observed and quantified in coherent high-frequency radar data
\item Analysis shows an increase in the likelihood of low elevation angle long-ducting Pedersen-mode under MSTID influence
\end{keypoints}

\begin{abstract}
Trans-ionospheric high frequency (HF: 3-30 MHz) response to gravity waves (GWs) is studied in the middle-latitude ionosphere in relation to thunderstorm activity. SuperDARN HF radar observations 
compared against the model simulations to quantify the impact of GW-generated MSTID (medium-scale traveling ionospheric disturbances) activity on the skywaves traveling through ionospheric F-region heights. The tropospheric thunderstorm-driven convective source is modeled using MAGIC. The outputs are coupled with GEMINI to model ionospheric plasma response, which is then used to model SuperDARN HF radar observations using the PHaRLAP raytracing tool. Semi-concentric GWs were observed at different atmospheric heights, creating MSTIDs at F-region heights. PHaRLAP raytracing through the modeled ionosphere shows great qualitative agreement with SuperDARN daytime ground scatter observations. Modeled rays show possibilities of long ducting Pedersen rays, suggesting MSTID can create a plasma waveguide to duct rays at the F-region height. 
\end{abstract}

\section*{Plain Language Summary}

%
%

%


%
%
%
%

\section{Introduction}
Atmospheric gravity waves (GWs) are mechanical oscillations driven by buoyancy restoring forces acting against gravity 
\cite<e.g.,>{hines1960,Francis1973,yeh1974acoustic,godin2012acoustic}, stemming from various sources, including convective storms, hurricanes, jet stream instabilities, topographic forcing, natural hazards (e.g., earthquakes, volcanic eruptions), human activity, and ocean-atmosphere interactions \cite<e.g.,>{nishioka2013concentric,azeem2015,plugonven2014,hickey2009propagation}. As GWs propagate vertically, their behavior is modulated by atmospheric stratification and background winds, \cite{fritts2003gravity,kshevetskii2021spectra} 
GWs often evolve nonlinearly, forming shocks or undergoing wave breaking \cite{heale2022primary,nozuka2024earthquake}. 
These vertically transported GWs often induce perturbations in the ionosphere ($\sim$80–500 km), producing plasma fluctuations known as traveling ionospheric disturbances (TIDs). To distinguish from other TID sources, we refer to these as GW-TIDs. 
Medium-scale traveling ionospheric disturbances (MSTIDs) density perturbations in the F-region ionosphere, characterized by periods of 15 to 60 minutes, wavelengths spanning several hundreds of kilometers, and horizontal propagation speeds ranging from 100 to 250 m/s \cite<e.g.,>{Ogawa1987}. These perturbations result from momentum and energy transfer between neutrals and plasma, with dynamics further shaped by geomagnetic and electrodynamic influences \cite{hines1960,hickey2009propagation}. In the lower ionosphere (D- and E- regions), GW-TIDs closely follow neutral motions, whereas at F-region altitudes, plasma responses are more aligned with magnetic field lines \cite{hooke1969region,schunk2000ionospheres}.

Medium-Scale Traveling Ionospheric Disturbances (MSTIDs) have been linked to both electrodynamic processes~\cite<e.g.,>{lin2022rapid, shinbori2022electromagnetic, Zhang2022GL100555} and atmospheric gravity waves (GWs). MSTIDs driven by electrodynamic enhancements typically occur at night and are observed as airglow brightenings. These are often associated with increased plasma-neutral coupling under specific geometrical conditions~\cite{ogawa2009medium, kelley2011origin}, which are beyond the scope of this study. MSTID impacts have been examined using both radio-based instruments (e.g., radar~\cite{}, ionosondes~\cite{}, and GPS Total Electron Content measurements~\cite{}) and optical techniques (e.g., all-sky imagers and Fabry–Perot interferometers)~\cite{}. Geomagnetic activity is also known to generate GWs via mechanisms such as Lorentz forcing and Joule heating, which in turn can influence over-the-horizon (OTH) high-frequency (HF: 3–30 MHz) radio communication~\cite{}. More recently, several studies have demonstrated the utility of OTH/HF radars and amateur radio networks in identifying MSTID signatures and their effects on HF propagation~\cite{Frissell2014, Frissell2016, Frissell2022}. While these studies have primarily focused on characterizing MSTIDs from various observational platforms, limited attention has been given to MSTIDs generated by GWs originating from convective sources, such as thunderstorms, and their specific impacts on HF communication. In this study, we address this gap by investigating the influence of GW-driven MSTIDs, particularly those generated during intense tropospheric events like thunderstorms, on trans-ionospheric HF communication channels.

In a vertically stratified ionosphere, high-frequency (HF) rays can travel long distances—up to 5000 km—via one-hop propagation, a mechanism commonly referred to as the Pedersen mode\cite<e.g.,>{Erukhimov01101997}. Pedersen rays are characterized by their divergent paths that glide along the peak of the F-region ionosphere (hmF$_2$)\cite{Ponomarenko2023RS007657}. Unlike low-elevation long-ducting modes, Pedersen rays exhibit distinct propagation characteristics, which depend on the HF operating frequency, ray elevation angle, and ground range\cite{davies1990ionospheric}. Due to their exponential divergence with distance, Pedersen rays are typically observed within a narrow elevation angle range at moderately high elevation angles under quiet ionospheric conditions~\cite{Ponomarenko2015}. Long-distance HF propagation in the Pedersen mode is thought to be supported by structured ionospheric features that confine the rays near the F-layer maximum. These guiding structures can result from GW-induced stratification~\cite{tinin1983propagation} or from large-scale, field-aligned irregularities formed along the layer~\cite{Tinin91RS02921}. Despite their potential importance, the role of horizontally tilted or stratified ionospheric structures—such as those induced by MSTIDs in reorganizing HF communication links has received limited attention. In this study, we present a modeling analysis demonstrating how MSTID-induced ionospheric plasma structures can enable and enhance Pedersen-mode propagation at relatively lower elevation angles than typically expected under undisturbed ionospheric conditions.

This study presents the first results from two-dimensional HF ray-tracing simulations through an MSTID-perturbed ionosphere, modeled using a three-dimensional atmosphere-ionosphere coupled framework. The simulations focus on a severe weather event that occurred over the continental United States (CONUS) in May 2017. Section~\ref{ds} details the observational datasets, modeling tools, and simulation configurations used in this event study. In Section~\ref{study}, we examine the observations, numerical model outputs, and the resulting data-model comparisons, with a focus on the influence of MSTIDs on Pedersen-mode propagation. Finally, Section~\ref{discussionandconclusion} summarizes key findings, discusses open challenges, and outlines future research directions in the context of past studies.

\section{Datasets \& Methodology}
\label{ds}
\subsection{SuperDARN}
\begin{figure}[ht!]
\centering
\includegraphics[width=1\textwidth]{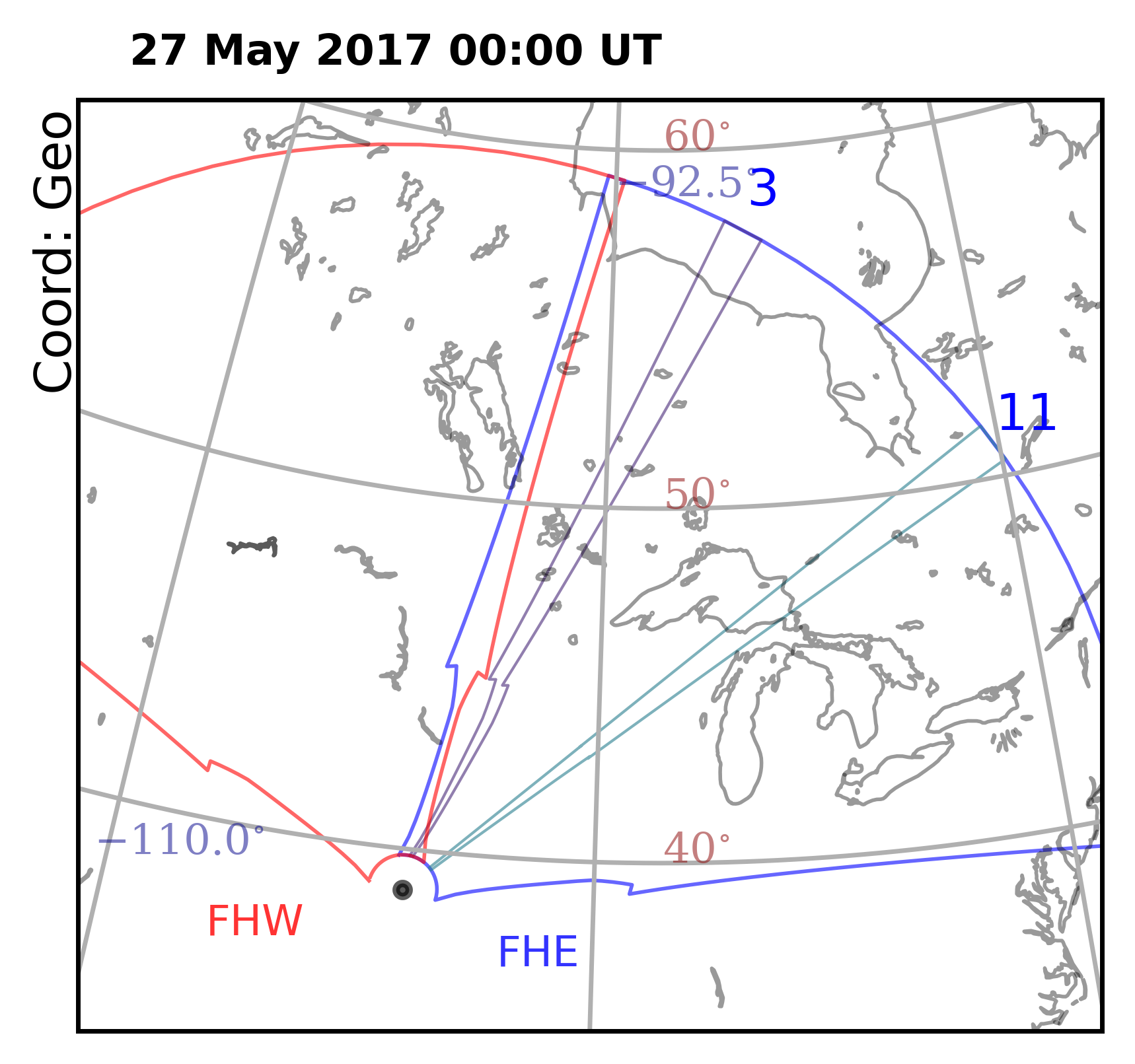}
\caption{Field-of-views (FoVs) of the SuperDARN Fort Hays East (in blue) and West (in red) radars (up to 50 range gates) located at middle latitude used in this study. Beams 3 and 11 of FHE radar are indicated by the colored contours and beam numbers mentioned at the end of the FOV in blue.} 
\label{fig:01}
\end{figure}

Super Dual Auroral Radar Network (SuperDARN) observations are the primary dataset used in this study to investigate the impacts of GW-generated MSTIDs on HF propagation. SuperDARN is an international collaboration of HF radars distributed across the middle, high, and polar latitudes of both hemispheres ~\cite{Chisham2007,Greenwald1985,Nishitani2019}. The radars generally operate between 8 and 18 MHz. In normal mode, each radar observes the line‐of‐sight (LoS) component of plasma velocity along 16 to 20 beams in 75-110 range gates spaced 45 km apart beginning at 180 km range. The typical integration time of each beam sounding is 3 or 6-seconds, which results in a full radar sweep through all beams in 1 or 2-minutes, respectively. Figure~\ref{fig:01} shows the location of the SuperDARN Fort Hays radars and their Field-of-Views (FoVs). Observations from these radars are used in this study.

\begin{figure}[htbp!]
\centering
\includegraphics[width=1\textwidth]{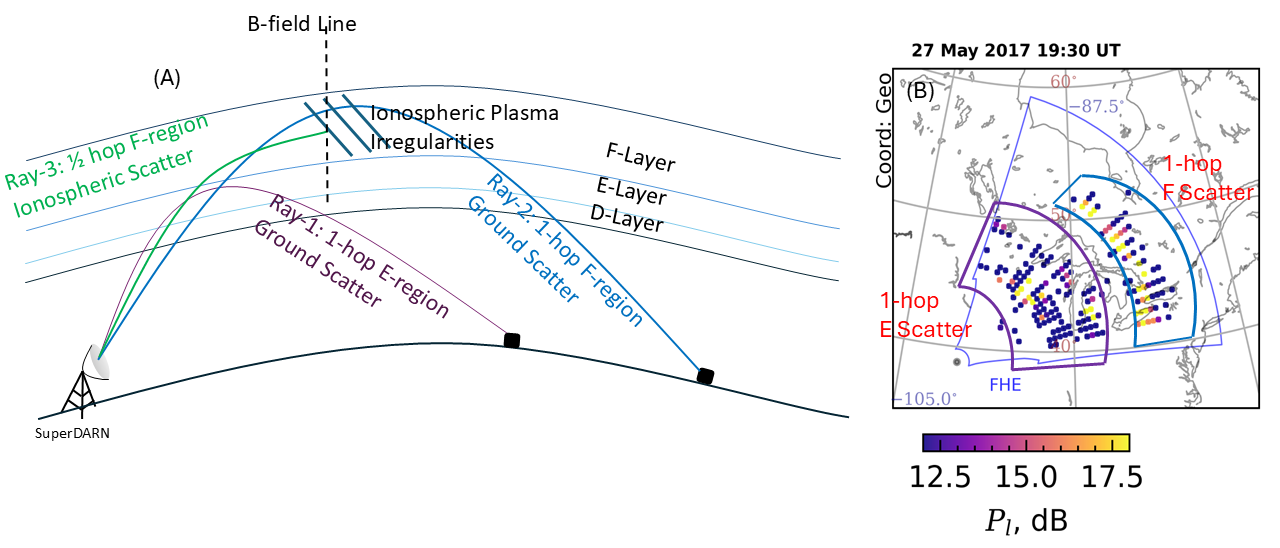}
\caption{(A) Schematic representation of SuperDARN radar ray paths illustrating ground scatter and ionospheric scatter. Illustration of 1/2-hop F-region, 1-hop E-region, and 1-hop F-region propagation geometries. The black squares indicate ground backscatter locations. (B) Field-of-view scan from the SuperDARN Blackstone radar, displaying line-of-sight backscatter returned power measurements taken on May 27, 2017, at 19:30 UT. The power is color-coded based on the scale provided at the bottom. Various scatter hops are identified and labeled within enclosed regions with red text.}
\label{fig:02}
\end{figure}

In this study, SuperDARN fitted parameter, specifically backscattered power from the Auto-Correlation Function (ACF~\cite{superdarn_data_analysis_working_group_2025_4435150}), is used to understand the impacts of MSTIDs on HF propagation. SuperDARN observations primarily consist of two types of backscatters, namely, ionospheric and ground scatters. Figure~\ref{fig:02}(A-B) illustrates the different propagation mode, scatter types, and provides an example of a backscattered power FoV plot from SuperDARN Fort Hays East radar.

In the case of ground scatter (GS), due to a high daytime vertical gradient in the ionospheric refractive index, the rays bend toward the ground and then are reflected from surface roughness, and return to the radar following the same paths. Based on the reflection height and return hops, we can further categorize the GS. Rays 1 and 2 in Figure~\ref{fig:02}(A) provide illustrative examples of daytime ground scatter, where Ray 1 and 2 represent rays reflected from E- and F-region altitudes, respectively. Figure~\ref{fig:02}(B) presents an example FoV fan plot showing backscattered power from daytime 1-hop E- and F-region ground scatters recorded by SuperDARN Fort Hays East radar. Along with backscatter power, GS `skip distance' can also provide valuable information and insight into the MSTIDs. The `skip distance' is defined as the first slant range where ground scatter is detected, is a key parameter that is considerably affected by ionospheric MSTIDs~\cite<e.g.,>{Frissell2014}. Ionospheric scatter (IS) occurs when a transmitted signal is reflected from ionospheric irregularities (refer to Ray 3 in Figure~\ref{fig:02}(A)). Previous studies used SuperDARN back-scattered return power of the GS observations to identify and characterize MSTIDs~\cite<e.g.,>{Frissell2014,Frissell2016}. In this study, we use the backscattered return power of the GS observations to investigate GW-generated MSTIDs over CONUS.



\subsection{Model Framework}
In this section, we describe the modeling framework that has been used to estimate the effects of GW-driven MSTIDs on HF radiocommunication. We performed simulations of 3D coupled atmosphere-ionosphere processes with MAGIC and GEMINI models, the results of which were then used in PHaRLAP HF ray-tracing model to find probable geolocation the HF rays with modified ionosphere. 
We shortly describe MAGIC and GEMINI simulation specifications and results and refer to the study by Inchin et al. (JGR, \textit{in review}, 2025), who discussed them in detail.

The generation and propagation of GWs were simulated with the three-dimensional, nonlinear, compressible Model for Acoustic-Gravity wave Interactions and Coupling (MAGIC) \cite{snively2003}. GW sources in the MAGIC model were approximated using latent heating profiles derived from Next Generation Weather Radar (NEXRAD) digital precipitation rate observations \cite{heale2019}. The atmospheric state was specified using the global empirical model NRLMSISE-00 \cite{picone2002} and temporally varying meridional and zonal winds were constructed by combining Modern-Era Retrospective Analysis for Research and Applications Version 2 (MERRA-2) data below 50 km with results from the HWM-14 empirical model for altitudes between 50 and 450 km \cite{drob2015}.


To simulate ionospheric plasma responses to GWs, we used the multi-species, nonlinear ionospheric model GEMINI \cite{zettergren2012ionospheric, zs2015}. The model was run using an open tilted dipole grid configuration to achieve high spatial resolution in the E- and F-regions of the ionosphere, with a grid resolution of $\sim$10 km. Three-dimensional outputs from the MAGIC simulation—including major neutral species densities (O$_2$, O, N$_2$), temperature perturbations, and horizontal and vertical fluid velocity perturbations—were used as external drivers for GEMINI, with a temporal resolution of 60 seconds. The thermospheric state and wind specifications in GEMINI were provided by the NRLMSIS-00 and HWM-14 empirical models, consistent with the setup used in MAGIC. Additionally, to investigate HF signal behavior in the absence of gravity waves, we performed a control GEMINI simulation that excluded GW drivers, hereafter referred to as the control run.

\subsubsection{PHaRLAP: HF Ray-Tracing Model}
We used Provision of High-Frequency Raytracing Laboratory for Propagation Studies (PHaRLAP) ray tracing model to geolocate the SuperDARN rays in the MSTID modified ionosphere~\cite{Cervera2014}. PHaRLAP incorporates a range of ray tracing engines with varying levels of sophistication, from 2D ray tracing to comprehensive 3D magnetoionic ray tracing. The 2D and 3D ray tracing modules are implementations of the equations developed by~\citeA{COLEMAN1997,Coleman1998} and \citeA{HASELGROVE1963}, respectively. The 2D ray tracing module utilizes ionospheric parameters, HF ray properties, and the elevation and bearing angles of the ray as inputs, generating the height and ground range of the traveling HF ray in kilometers as output. In contrast, the 3D ray tracing module outputs the height, latitude, and longitude of the traveling HF ray. For our research, we employed the 2D ray tracing module to geolocate rays along each individual beam of the SuperDARN Fort Hays radar.

\subsubsection{Estimating Modeled Backscattered Power}
In this study, we were primarily focused on analyzing and comparing SuperDARN ground scatter observations. Output from PHaRLAP was used to estimate the backscattered power.

To estimate the modeled ground-scattered power, we adopted the methodology outlined in ~\cite{de_larquier_mid_latitude_2013}. First, simulated rays that reach the ground after reflecting in the ionosphere were identified as GS. These rays are then grouped into 45 km slant-range bins, and the number of rays in each bin is counted. The counts are subsequently weighted by a factor of $\frac{1}{r^3}$, where $r$ is the slant range. This factor accounts for the geometric power decay, assuming the target’s size increases linearly with range, as is typical for GS. The choice of 45 km range bins are considered to SuperDARN operational range gates. Note that this model does not include the terrain geometry or the reflective properties of the target, which may be separately investigated in the future.

\section{Event Analysis: 27 May 2017}
In this section, we present a GW-generated ionospheric MSTID event stemming from tropospheric thunderstorm-driven convective source events on 27 May 2017. These MSTIDs were observed by the middle latitude Fort Hays radars. We then describe the output of simulated rays from the model framework and compare it with radar observations. We also analyze the influence of MSTIDs on special long ducting HF mode, also referred to as Pedersen mode.

\begin{figure}[h!]
\centering
\includegraphics[width=1\textwidth]{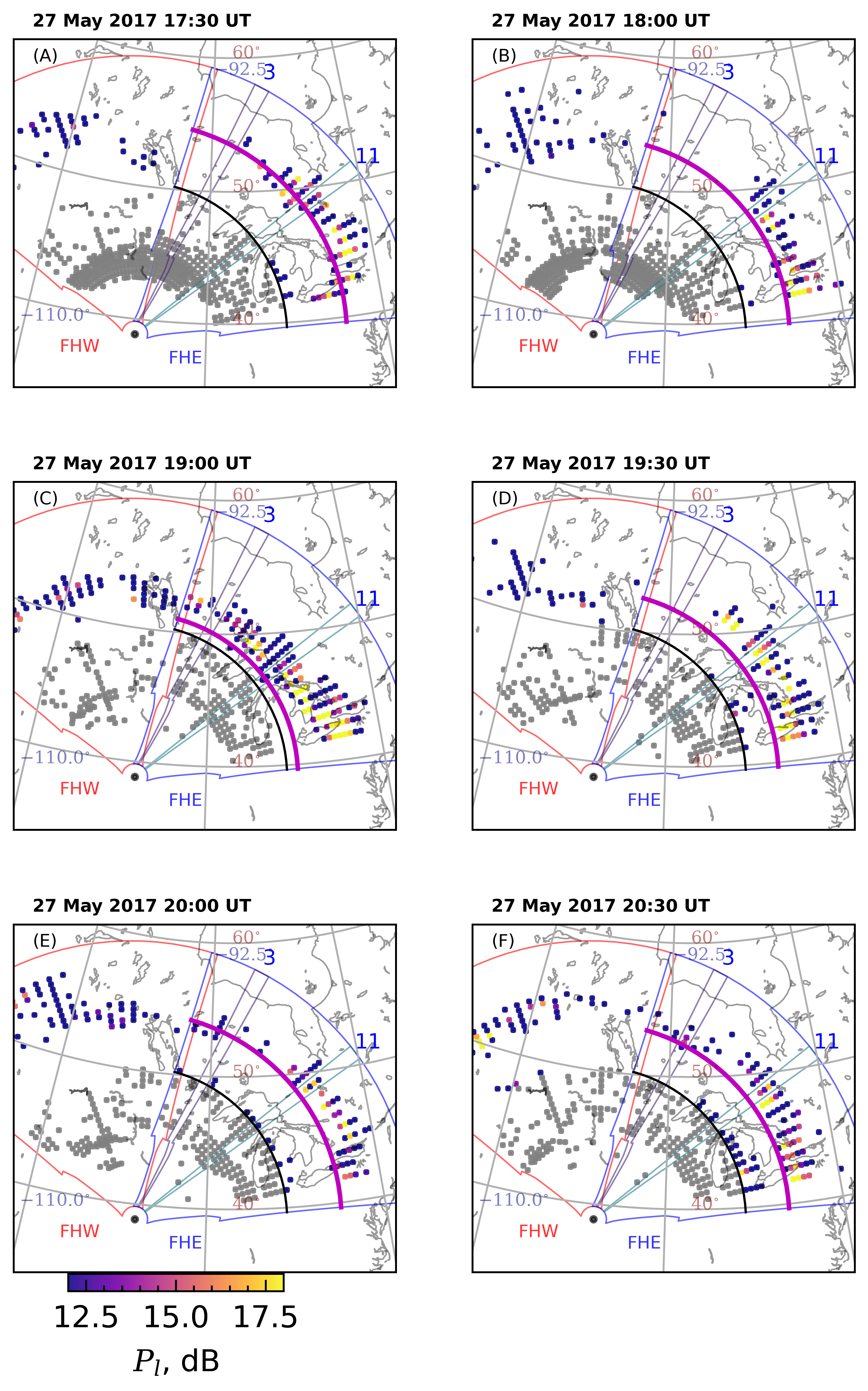}
\caption{Fields-of-View (FOV) scan plots from the SuperDARN Fort Hays radars (FHE and FHW) during GW-generated MSTIDs on May 27, 2017:
Panels (A–E) present a series of FOV scan plots depicting backscattered power, color-coded according to the scale shown in the bottom-left corner. Beams 3 and 11 of the FHE radar are highlighted with colored contours within the larger FOV outlines and are also labeled in blue text. The green solid curve overlaid on the FHE FOV in each panel marks the estimated skip distance of the furthest ground-scatter returns. The black solid curve serves as a visual reference, indicating the extent of the most distant ground-scatter population.}
\label{fig:03}
\end{figure}

Figure~\ref{fig:03} illustrates the impact of MSTIDs on observations from the SuperDARN HF radars. The figure showcases radar FoV scan plots from the SuperDARN Fort Hays (FH) radars, which are strategically located in central Kansas and have FoV covering a significant portion of central America. Each panel in Figure~\ref{fig:03} displays the line-of-sight (LoS) backscattered fitted power ($P_l$) obtained from a 1-minute radar scan, with the panels presented at 30-minute intervals to highlight temporal variations. The backscattered power is color-coded according to the color bar on the bottom left corner. The figure contains six panels, labeled (A) through (F), each displaying a map of backscatter power from FH radars. The time of each scan is shown at the top of each panel, beginning at 18:00 UT and ending at 20:30 UT, with each scan separated by 30 minutes. Observations from both the radars reveal how the MSTIDs, driven by convective atmospheric sources, modulate ionospheric conditions, affecting the radar backscatter patterns, specifically skip distances. For ease of understanding, we are going to refer to and analyze the observation of FHE radar with MSTID signatures. We see two populations of backscatters within the FoV of FHE radar, near (within 30 range gates, $\leq$1300 km) and far ($\geq$1500 km) from the radar station. The near-range scatters are grayed out and are not of interest for this study. The backscatter population far from the radar station is ground scatter. The black and magenta curved lines mark the reference and skip distance of the backscatter population at far distances. The skip distance of the ground scatter population recorded a significant motion (motion towards and away from the radar station) over the 3-hour window. Additionally, we see variations in intensification and lessening of backscatter power values across the beams and with time, indicating ionospheric density compression and rarefactions caused by MSTIDs passing through the FOVs of the radars~\cite{Frissell2014}. 

\begin{figure}
    \centering
    \includegraphics[width=1.2\linewidth]{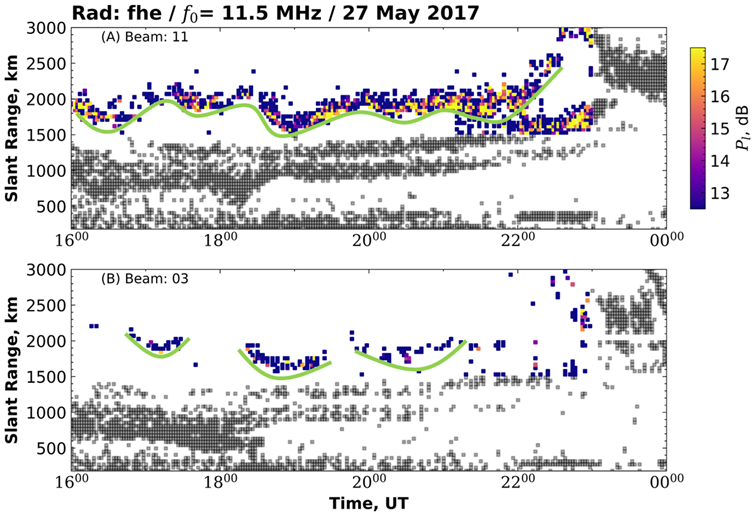}
    \caption{Observations from the SuperDARN Fort Hays East radars along beams (A) 11 and (B) 03: Range-Time-Intensity (RTI) plots of the backscattered power showing HF response to ionospheric MSTIDs on May 27, 2017, color coded by the color bar on the right. The green curves in both panels indicate the tentative location of the skip distance of the furthest ground-scatter population, indicating the MSTID impacts on the transitionospheric HF communication. The grayed scatter are not of interest for this study.}
    \label{fig:04}
\end{figure}

We also use a different data plotting format, namely range-time-intensity (RTI) plot, to observe motion in skip distance locations along beams 3 and 11 of the FHE SuperDARN radar. Figure~\ref{fig:04} presents backscatter power ($P_l$) along beams 3 and 11 of the FHE radar, as a function of time (horizontal axis) and slant range (vertical axis). The backscatter power is color-coded by color bar on the right of panel (A). The geographic region along these two beams can be referred back to Figure~\ref{fig:03}. We see different backscatter populations located at different times and spaces, indicating changes in the ionospheric propagation conditions. We are specifically interested in ground scatter populations between slant range 1500-2000 km and 16-21 UT. We have added the green curves in both panels, indicating the motion in skip distance of the ground scatter population located between 1500-2500 km slant ranges. Similar to FoV plots (Figure~\ref{fig:04}), we see motion in the skip distance. The grayed scatters are not of interest for this study. We see a few notable features in this RTI plot: (1) skip distance of the ground scatter population recorded a significant motion; (2) variations in intensification and lessening of backscatter power values indicating ionospheric density compression and rarefactions. We speculate the density rarefaction effects are more intense along beam 3 than along beam 11, creating total loss-of-signal between the three different GS populations indicated by green curves. The MSTID activity is quite complex; multiple, different MSTID wave trains can be seen in the RTI plots by observing changes in skip distance location.

\subsection{Data-Model Comparison}
\label{study}
We simulated atmosphere-ionosphere coupled processes related to the May 27th, 2017, severe weather system (WS) episode over midwestern United States. This event was chosen due to a geomagnetically calm ionosphere on the day of interest and prior to this date, but also a small presence of TIDs from other sources, whereas TIDs from the WS had distinctive concentric patterns as observed in GNSS TEC. 
As the most prominent TIDs were observed between 18-21 UT of 05/27/2017 (see the animation in the Supporting Information, SI), we constrained our simulations by 16-22 UT time window to accommodate available computational resources. The event occurred on a geomagnetically calm day, with average daily Dst, Ap, and F10.7 indices of 7 nT, 13 nT, and 84.1, respectively (retrieved from NASA OmniWeb, \cite{king2005}).

\begin{figure}[h!]
\centering
\includegraphics[width=1\textwidth]{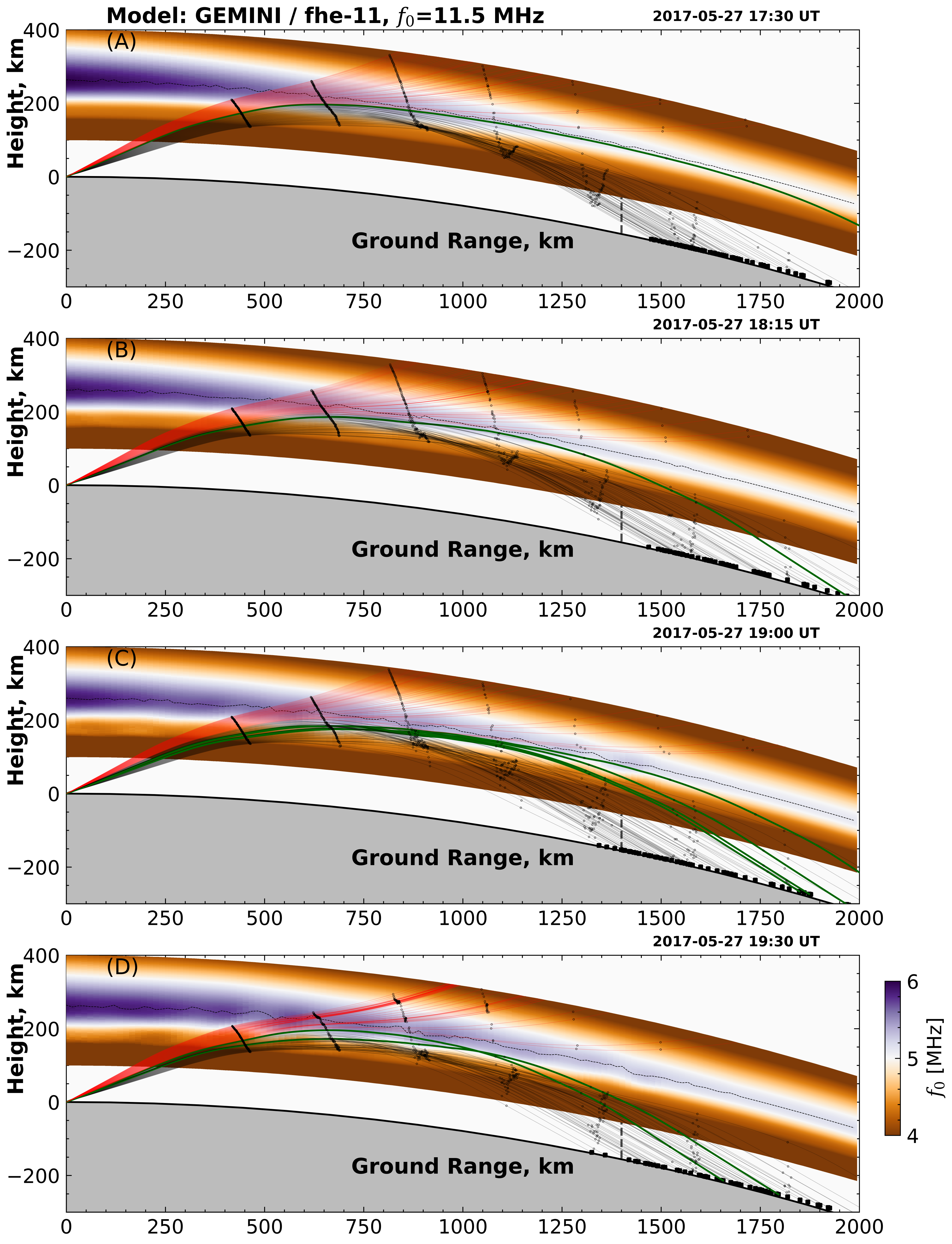}
\caption{PHaRLAP Raytracing Through MSTID-Modified (MAGIC+GEMINI) Ionosphere on May 27, 2017: Simulated rays from the SuperDARN Fort Hays East (FHE) radar along beam 11 are shown with elevation angles ranging from 18$^\circ$ to 30$^\circ$ at four time instances: (A) 17:30 UT, (B) 18:15 UT, (C) 19:00 UT, and (D) 19:30 UT. The radar was operating at 11.5 MHz. Escaping, long ducting Pedersen mode, and ground rays are indicated by red, thick green, and black curves, respectively. The vertical dashed line at 1400 km serves as a reference to illustrate the back-and-forth motion of the radar skip distance in response to the phase fronts of the MSTID. Black squares on the ground indicate the ground-scatter (GS) reflection points. MAGIC+GEMINI simulated plasma frequencies are overlaid on each panel and color-coded according to the scale shown in the bottom right panel. The dashed back curve indicates the hmF$_2$. The gray shaded region shows the Earth and its curvature. Black contours indicated by black dots represent the group range between 500-2000 km with a separation of 250 km.}
\label{fig:06}
\end{figure}

Figure~\ref{fig:06} shows output from PHaRLAP raytrace through MAGIC+GEMINI simulated ionosphere along the beam 11 of FHE radar (refer to Figure~\ref{fig:01} to check the geographic location), which was operating at 11.5 MHz. Panels (A-D) present outputs from four different timestamps, 17:30 UT, 18:15 UT, 19 UT, and 19:30 UT, respectively. Elevation angle of the simulated rays presented in this figure ranges from 18$^\circ$ to 30$^\circ$. Simulation suggested possible escaping rays (in red), long ducting Pedersen mode rays (in green), and rays reaching the ground (ground scatters), respectively. The black squares on the ground indicate the reflection point of the modeled GS. We overlaid the simulated ionospheric plasma density from MAGIC+GEMINI, and color-coded it by the color bar on the bottom left corner. The dashed back curve indicates the hmF$_2$ location. The gray shaded region shows the Earth and its curvature (by scale). Contours indicated by black dots represent the group range between 500-2000 km with a separation of 250 km. The vertical line at 1400 km on ground serves as a reference line to illustrate the back-and-forth motion of the radar skip distance in response to the phase fronts of the MSTID. Simulated MSTID can be observed in the field of view of the radar beam around panel (C-D), which alters the skip distance of the ground scatter rays. We see a significant fluctuations in the higher group ranges in response to MSTID. We see MSTID altered the propagation condition of the ionosphere creating possibilities of long ducting Pedersen modes at other elevation angles, which is also a function of phase fronts. Simulation shows that the MSTID alters the rays passing through the ionosphere by compression and rearfraction of the plasma density by creating medium-scale ($\sim$100-1000 km) plasma density structures.

\begin{figure}[h!]
\centering
\includegraphics[width=1\textwidth]{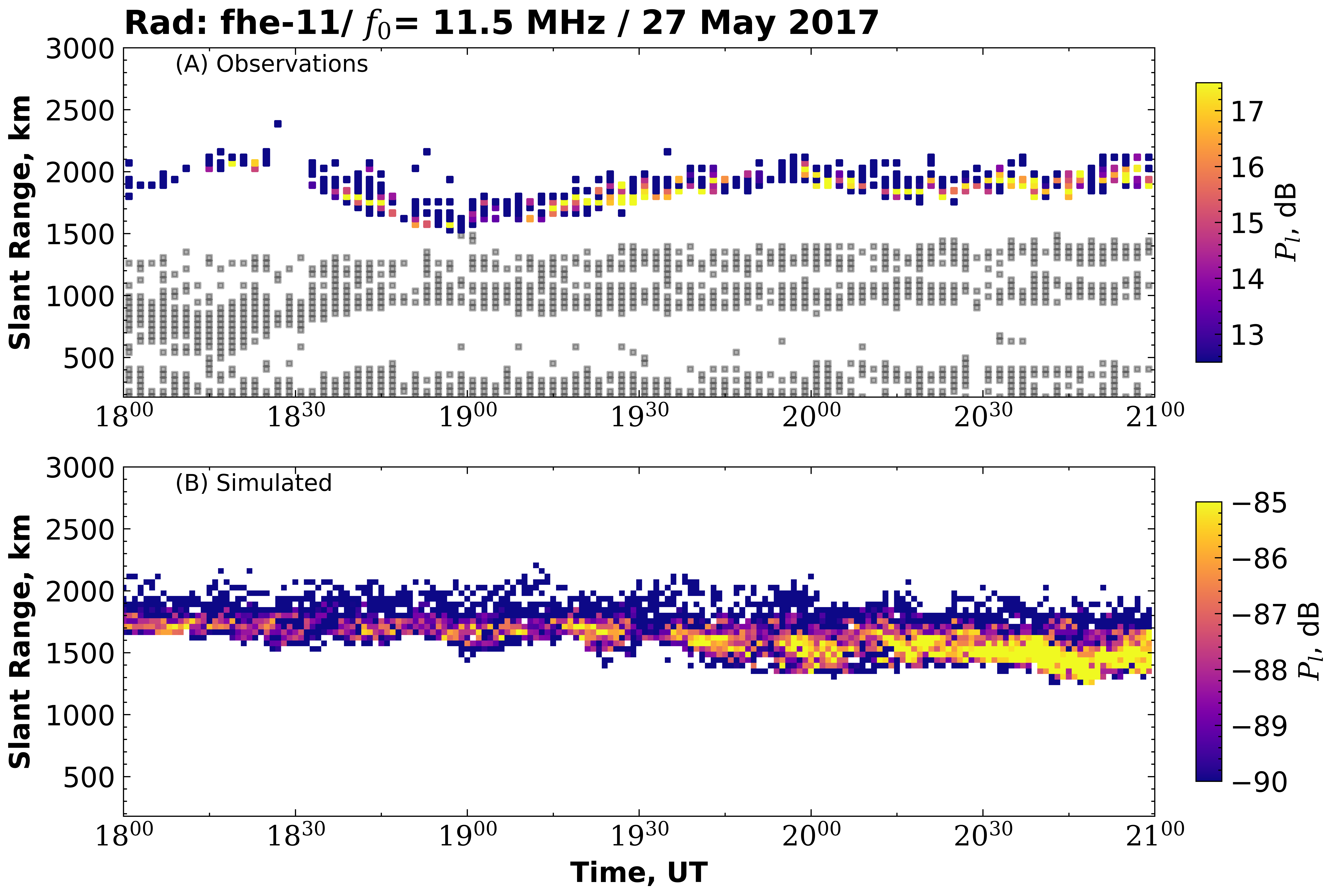}
\caption{Range-Time-Intensity (RTI) plots are presented, comparing (A) observed and (B) simulated ground scatter from the SuperDARN FHE radar (beam 11). These plots illustrate a data-model comparison highlighting the impact of MSTIDs on the radar-observed ground scatter characteristics. In both panels, observed and simulated power levels are color-coded according to the respective color bars on the right. Backscatter data within the initial 1500 km slant range are masked (grayed out) as they fall outside the scope of this particular study. The depicted oscillations in ground scatter skip distances, along with concurrent fluctuations in received power, demonstrate the influence of MSTIDs on transionospheric HF radio wave propagation, which directly affects communication channels.}
\label{fig:07}
\end{figure}

Figure~\ref{fig:07} presents the RTI plots showing observed and simulated ground scatter fitted power ($P_l$) and relative power ($P_r$), respectively, from FHE radar along beam 11. Section 2.2.2 describes the methodology to compute $P_r$ from PHaRLAP simulated HF rays. Note that the relative power is not an exact representation of the backscatter power received by the SuperDARN radar, hence, we will not be able to conduct regression error analysis between the observed and simulated power. We, however, think it is possible to conduct a feature-based comparison between observations and simulated outputs.

Figure~\ref{fig:08} presents the FoV scan plots of backscattered relative modeled power ($P_r$) from FHE radar, with the panels presented at 30-minute intervals to highlight temporal variations. The backscattered power is color-coded according to the color bar on the bottom left corner. The time of each scan is displayed at the top of each panel, beginning at 18:00 UT and ending at 20:30 UT, with each scan separated by 30 minutes. The black and green curved lines mark the reference and skip distance of the backscatter population. Additionally, we overplotted the skip distance (magenta curve) from observation (refer to Figure~\ref{fig:03}). Note that the skip distance of the modeled ground scatter population recorded motion (motion towards and away from the radar station) over a 3-hour window presented in this figure. Additionally, we see variations in intensification and a lessening of backscatter power values across the beams and with time, indicating ionospheric density compression and rarefactions caused by MSTIDs passing through the FOVs of the radar. This shows a reasonable agreement between data and model outputs.

\subsection{Special long ducting HF propagation mode: Pedersen mode}
While MSTID-driven oscillations in ionospheric electron density affect the skip distance by perturbing the trajectory of transionospheric HF rays, our simulations also reveal the potential formation of long-distance ducted propagation modes, commonly referred to as Pedersen modes. These modes occur within a narrow angular range of elevation and are characterized by divergent rays that glide along the ionospheric peak (hmF$_2$)~\cite{Ponomarenko2015}. In an unperturbed ionosphere, such Pedersen modes are expected to exist only within a narrow communication channel, determined by specific combinations of elevation angle and frequency. However, our simulations suggest that in a TID-perturbed ionosphere, multiple such narrow channels may emerge, each capable of supporting Pedersen-like ducted propagation (see Figure~\ref{fig:06}). While no observational evidence of Pedersen modes was detected by the Fort Hays radars in this particular case, the simulations suggest that MSTID-induced perturbations may create conditions favorable for their occurrence.

\begin{figure}[h!]
\centering
\includegraphics[width=1\textwidth]{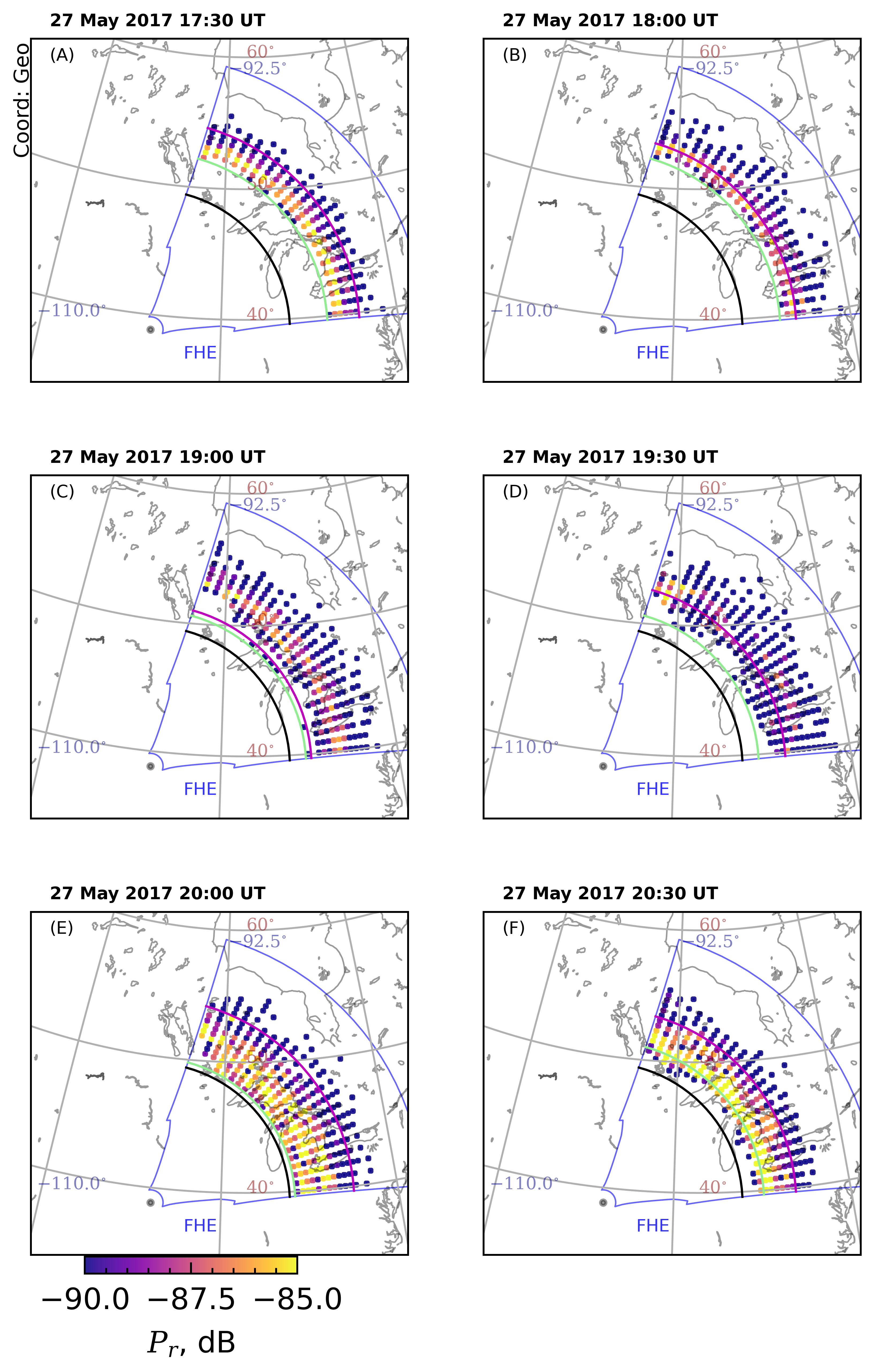}
\caption{Fields-of-view (FoV) scan plots of the backscattered from the SuperDARN FHE radar simulated using MAGIC+GEMINI and raytraced using PHaRLAP. Panels (A-E) present a series of FOV scan plots depicting relative backscattered power, color-coded according to the scale shown in the bottom-left corner. The magenta and green solid curves overlaid on the FHE FOV in each panel mark the estimated skip distance of the furthest simulated and observed ground-scatter returns, respectively. The black solid curve serves as a visual reference, indicating the extent of the most distant ground-scatter population.}
\label{fig:08}
\end{figure}

To investigate this further, we present a focused analysis of modeled Pedersen-mode propagation in MSTID-perturbed conditions using Figure~\ref{fig:08}. We compare a narrow elevation angle range (21.7$^\circ$–22.9$^\circ$) at 11.5 MHz between a TID-perturbed ionosphere (Panel A) and a control simulation (Panel B) on 19:30 UT. Three rays were launched within this range at 0.1$^\circ$ separation. In the control case, no ducting or Pedersen mode is observed. However, in the presence of TIDs, a narrow communication channel emerges around 21.8$^\circ$, supporting a ducted Pedersen mode. Panel A of Figure~\ref{fig:08} shows a zoomed-in view of the three rays, highlighting the interaction between the MSTID structures and ray trajectories. All rays are influenced by the TID phase fronts, as evident from their modified skip distances compared to the control run. Of particular interest is the green-colored ray, which exhibits a gliding trajectory aligned with the ionospheric peak (hmF$_2$), shown by the black dashed curve. The top-right inset in Panel A illustrates how the TID structure facilitates upward bending and rotation of the ray, enabling it to propagate parallel to the ionospheric maximum over a ground range exceeding 250 km. This behavior indicates that the MSTID crest contributes to localized density refraction, allowing rays at lower elevation angles to reach higher altitudes and transition into Pedersen mode propagation.

\begin{figure}[h!]
\centering
\includegraphics[width=1\textwidth]{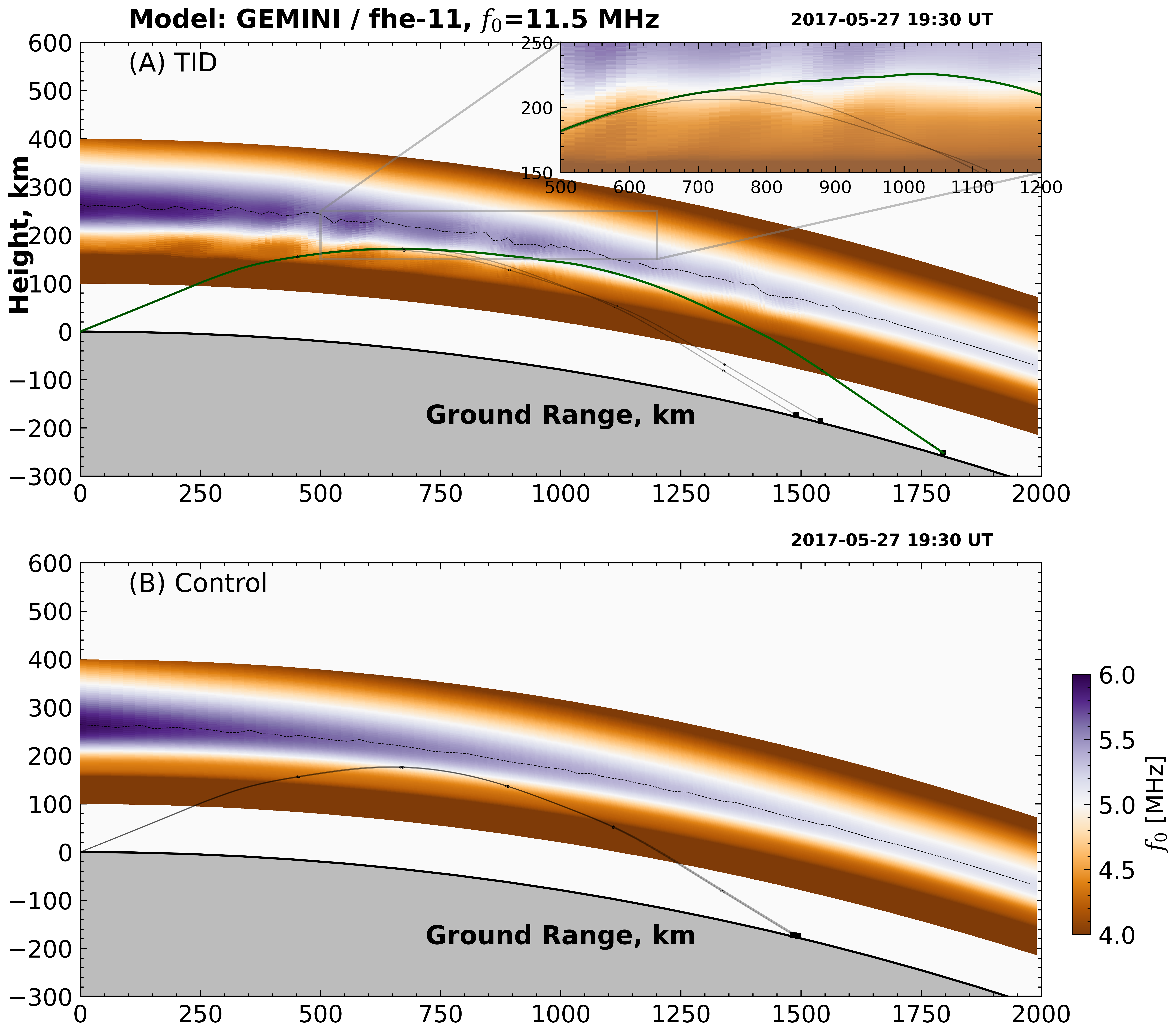}
\caption{PHaRLAP raytracing through (A) MSTID-modified, and (B) controlled ionosphere showing impact of MSTIDs on long-ducting HF modes. Simulated rays from the SuperDARN Fort Hays East (FHE) radar along beam 11 are shown with elevation angles ranging from 21.7$^\circ$ to 22.9$^\circ$ at 19:30 UT. The radar was operating at 11.5 MHz. MAGIC+GEMINI simulated plasma frequencies are overlaid on each panel and color-coded according to the scale shown in the bottom right panel.}
\label{fig:09}
\end{figure}

\begin{figure}[h!]
\centering
\includegraphics[width=1\textwidth]{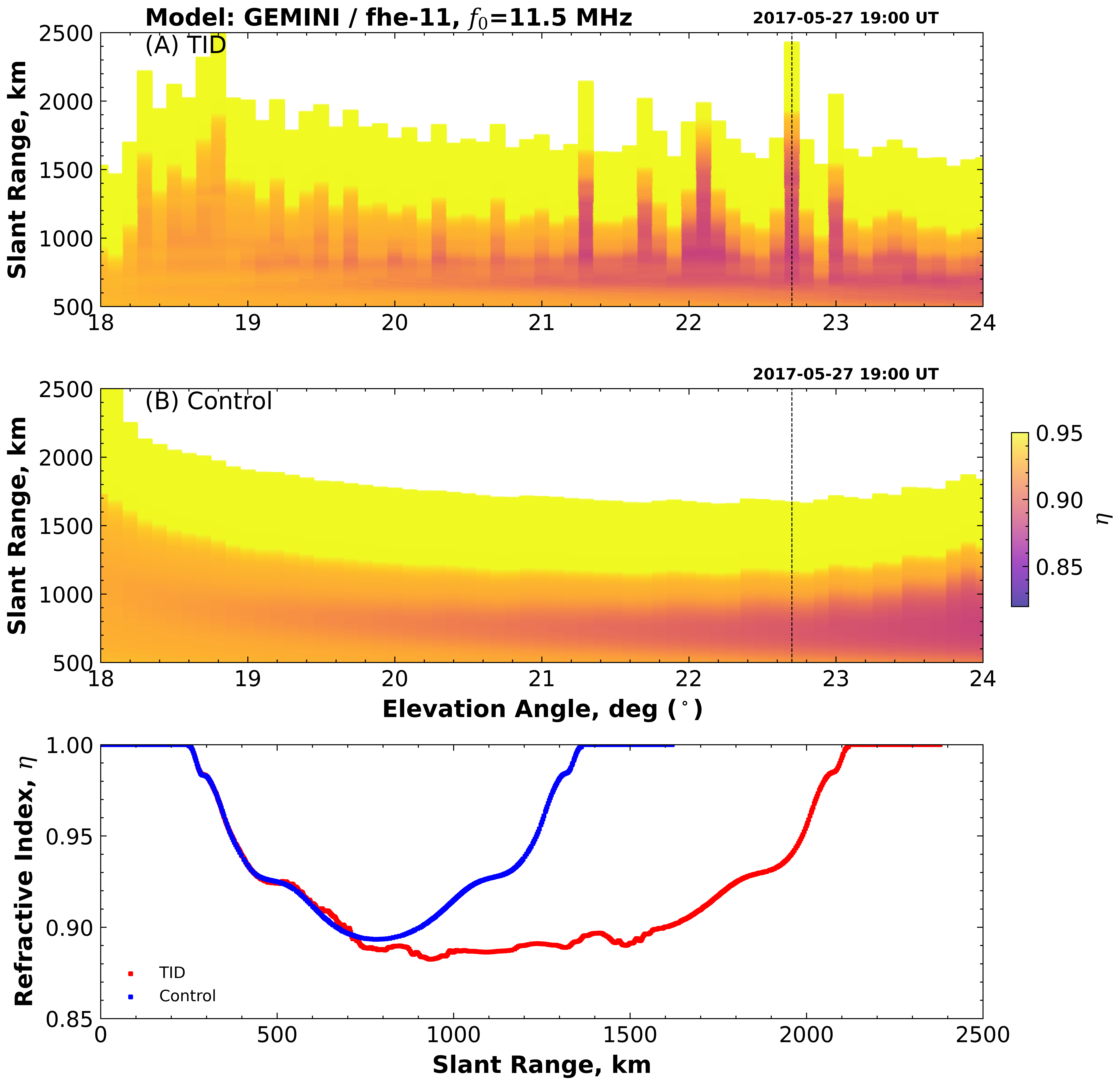}
\caption{The figure illustrates the simulated distribution of refractive index ($\eta$) as a function of slant range and elevation angle at 19:00 UT along beam 11 of the FHE radar operating at 11.5 MHz. Panels (A) and (B) compare propagation through an MSTID-modified ionosphere and a controlled (unperturbed) ionosphere, respectively, thereby demonstrating the impact of MSTIDs on long-range HF ducting modes. In both panels (A) and (B), the refractive index ($\eta$) is color-coded according to the scale shown by the color bar adjacent to panel (B). Panel (C) displays the refractive index profile as a function of slant range along a specific elevation angle of 22.8$^\circ$. This panel highlights how MSTID-induced fluctuations in the refractive index can create conditions favorable for the formation of a long-range Pedersen ducting mode.}
\label{fig:10}
\end{figure}

We further analyzed the occurrence of Pedersen modes by comparing ray trajectories simulated between elevation angles of 18$^\circ$ and 24$^\circ$ at 19:00 UT, as shown in Figure~\ref{fig:10}. Panels A and B present the refractive index $\eta$, plotted as a function of slant range and elevation angle. These refractive index distributions are derived from GEMINI simulations with and without MSTID (Control). Panel C of Figure~\ref{fig:10} shows the refractive index $\eta$ along the path of a selected ray at an elevation angle of 22.7$^\circ$, under both MSTID-perturbed (blue) and control (red) conditions. Several key observations emerge from this figure: 
\begin{itemize}
    \item Long ducting modes appear at lower elevation angles (specifically between 21° and 23°) in the MSTID simulation.
    \item Rays in the MSTID case traverse alternating regions of low and high plasma density, creating favorable propagation conditions for long-distance ducting at lower elevation angles.
    \item Localized MSTID structures (see Figure~\ref{fig:09}(A)) act as waveguides by steering rays through regions of low refractive index toward the ionospheric peak (hmF$_2$), enabling extended gliding and longer skip distances and creating HF conditions suitable for supporting long-ducting Pedersen mode.
\end{itemize}

\section{Discussions and Summary}
\label{discussionandconclusion}

This study investigates the impact of gravity wave (GW)-driven medium-scale traveling ionospheric disturbances (MSTIDs), generated by a severe thunderstorm over the Midwest and eastern United States on May 27, 2017, on high-frequency (HF) communication channels using SuperDARN radar observations and first-principles modeling. 3D MAGIC model, driven by NEXRAD-derived precipitation rates, was used to compute temporally and spatially varying latent heating profiles as GW sources~\cite{}. These outputs were then used to drive the GEMINI 3D multi-fluid nonlinear ionospheric model to simulate ionospheric plasma responses~\cite{}. The coupled MAGIC-GEMINI results were validated against GNSS vTEC and COSMIC-2 radio occultation observations and presented in~\citeA{}, demonstrating good agreement in MSTID periods ($\sim$5–25 min) and horizontal wavelengths ($\sim$100–300 km). Previous studies showed that the simulated vertical TEC fluctuations ($\sim$0.02–0.35 TECU) matched observations within an uncertainty of $\sim$15\%, suggesting high model accuracy for the event under consideration. These results provide a robust basis for conducting 2D PHaRLAP ray-tracing simulations to examine MSTID impacts on HF propagation observed by SuperDARN radars.

Both observations and model results reveal that gravity wave (GW)-driven medium-scale traveling ionospheric disturbances (MSTIDs) significantly affect HF communication channels. Observational data show notable fluctuations in ground scatter (GS) skip distances, leading to asymmetric backscattered power distributions across both time and space, as evident in SuperDARN RTI and FoV plots. The modeling framework successfully reproduces these MSTID-driven signatures in HF data, demonstrating its utility for studying convective source-driven ionospheric perturbations. The observed spatial asymmetry in signal power across radar beams indicates horizontally structured TID fronts, while temporal asymmetries reflect plasma density fluctuations caused by compression and rarefaction effects. These features are consistent with prior studies on MSTID impacts using HF observations~\cite{Frissell2014, Prikryl2022, Frissell2022}, which predominantly focused on auroral, magnetospheric, or volcanic sources~\cite{Kozlovsky2025, Zhang2022GL100555}. In contrast, our study emphasizes that convective tropospheric sources, such as severe thunderstorms, can also generate GWs capable of significantly disrupting HF propagation.

We explore how the presence of MSTIDs modulates the behavior of Pedersen mode propagation. In the absence of disturbances, as shown in Figure~\ref{fig:06}(A), the propagation characteristics of Pedersen rays are primarily governed by the electron density distribution near the peak of the F-layer. Prior simulation work by \citeA{eckermann1999global} demonstrated that when the F1 and F2 layers are closely spaced, the resulting flattened density profile minimizes energy leakage, thereby creating favorable conditions for sustained Pedersen mode propagation.

Our modeling results indicate that MSTID-induced ionospheric disturbances can enhance the likelihood of Pedersen modes emerging at lower elevation angles. This suggests that both structured and irregular plasma density features introduced by MSTIDs play a critical role in supporting these long-ducting propagation paths. Consistent with the findings of \citeA{eckermann1999global}, our simulations suggest that irregularities near the F-layer peak act as stabilizing features for Pedersen mode propagation during disturbed periods. However, their modeling also emphasizes that for isotropic irregular structures located symmetrically around the F-layer maximum, within a narrow vertical extent (~10–20 km), the formation of effective ducting channels remains unlikely.

Our future studies will focus on evaluating the impacts of MSTID-driven long-ducting Pedersen modes and how MSTID phase fronts (i.e., ionospheric tilts) influence HF communication. To do this, we will employ PHaRLAP’s 3D ray-tracing capabilities to examine the three-dimensional structure of ionospheric tilts. Additionally, we aim to assess the effectiveness of existing HF sounding techniques in capturing MSTID characteristics. Although HF sounder observations are limited in both spatial coverage and temporal resolution, model-guided analysis will allow us to evaluate the accuracy of key ionospheric parameters, such as NmF$_2$ and hmF$_2$—derived from instruments like Digisonde and VIPIR. This integrated approach will help identify which sounding techniques best capture the dynamic behavior of the ionosphere, especially during MSTID events.


\section{Open Research}
GNSS signal observations used for the investigation of MSTIDs in the study are available at (1) Scripps Orbit and Permanent Array Center via \url{http://sopac-csrc.ucsd.edu/index.php/gnss/} with public access, (2) The National Aeronautics and Space Administration the Crustal Dynamics Data Information System \cite{NOLL20101421} via \url{https://cddis.nasa.gov/archive/gnss/data/daily/} with public access, (3) National Oceanic and Atmospheric Administration’s National Geodetic Survey NGS via \url{https://geodesy.noaa.gov/CORS/} with public access and GAGE Facility operated by UNAVCO, Inc. via \url{https://www.unavco.org/data/gps-gnss/gps-gnss.html} with public access. GAGE Facility, operated by UNAVCO, Inc., is supported by the National Science Foundation, the National Aeronautics and Space Administration, and the U.S. Geological Survey under NSF Cooperative Agreement EAR-1724794. All the data and Python code are uploaded into the Zenodo available for public use~\cite{c22}. The analysis and visualization were completed with the help of free, open-source software tools such as matplotlib~\cite{Hunter2007}, IPython~\cite{pg07}, pandas~\cite{m10}, pyDARN~\cite{Shi2022}, and others~\cite<e.g.,>{ma11}. The SuperDARN data is available via \url{http://vt.superdarn.org/}. 

\acknowledgments
This research is supported by NASA LWS 80NSSC22K1022. The authors acknowledge the use of SuperDARN data. SuperDARN is a collection of radars funded by national scientific funding agencies of Australia, Canada, China, France, Italy, Japan, Norway, South Africa, United Kingdom, and the United States of America.
\bibliography{draft1} 
%




%
%
%
%
%

\clearpage

\end{document}